\documentclass[submission,copyright,creativecommons]{eptcs}
\providecommand{\event}{The 21st Workshop on Logic-based Methods\\ in Programming Environments (WLPE 2011)} 
\usepackage{breakurl}             
\usepackage{xcolor}
\usepackage{graphicx}

\title{A Well-typed Lightweight Situation Calculus\thanks{This work is also offered to present at \textit{the 20th International Workshop on Functional and (Constraint) Logic Programming (WFLP'11)}, Odense, Denmark, July 2011.}}
\author{Li Tan
\institute{Department of Computer Science and Engineering\\
University of California, Riverside\\
Riverside, CA, USA 92507}
\email{ltan003@cs.ucr.edu}
}
\def\titlerunning{A Well-typed Lightweight Situation Calculus}
\def\authorrunning{Li Tan}
\begin{document}
\maketitle

\begin{abstract}
Situation calculus has been widely applied in Artificial Intelligence related fields. This formalism is considered as a dialect of logic programming language and mostly used in dynamic domain modeling. However, type systems are hardly deployed in situation calculus in the literature.
To achieve a correct and sound typed program written in situation calculus, adding typing elements into the current situation calculus will be quite helpful. In this paper, we propose to add more typing mechanisms to the current version of situation calculus, especially for three basic elements in situation calculus: situations, actions and objects, and then perform rigid type checking for existing situation calculus programs to find out the well-typed and ill-typed ones.
In this way, type correctness and soundness in situation calculus programs can be guaranteed by type checking based on our type system.
This modified version of a lightweight situation calculus is proved to be a robust and well-typed system.
\end{abstract}

\section{Introduction}

Introduced by John McCarthy in 1963 \cite{SC}, situation calculus has been widely applied in Artificial Intelligence related research areas and other fields. This formalism is considered as a dialect of logic programming language and mostly used in dynamic domain modeling. Based on First Order Logic (FOL) \cite{FOL} and Basic Action Theory \cite{BAT}, situation calculus can be used for reasoning efficiently by virtue of dynamic elements, such as actions and fluents. Basic concepts of situation calculus are fundamentals of First Order Logic and Set Theory in Mathematical Logic, which greatly facilitate the process of action-based reasoning in situation calculus.

In order to make programs sound and correct in semantics, people have proposed type systems \cite{TAPL} to ensure such significant properties. A well-typed programming language is determined by two semantic properties: preservation and progress. The first property makes sure that types are invariant under the evaluation and typing rules. And the progress property says a well-typed program never gets stuck. Nevertheless, little attention has been put on equipping formal languages good at dynamic modeling and reasoning, like situation calculus, with strong typing mechanisms. Indeed, situation Calculus is a typed second-order formal language, but from the viewpoint of type checking, it is not enough to finish smoothly. For instance, in situation calculus, only typed quantifiers are introduced for basic variables, while as for other logical expressions consisting of variables and connectives, fluents and predicates, current situation calculus emphasizes little on how to type check whether they are well-typed, how to type them thoroughly. Therefore, equipping other elements in current version of situation calculus with types is greatly needed for a complete and robust programming language with its type system, which is definitely feasible according to our investigation.

In this paper, in addition to the handy available typed variables, we propose to add more typing mechanisms to three basic elements in situation calculus: situations, actions and objects, consider a classical scenario for a piece of program based on the modified lightweight situation calculus, and then perform rigid type checking for the situation calculus program. If type errors are found, we would provide corresponding recommendation on how to correct the program into a well-typed one. Furthermore, to support our ideas in practice, we implement a type checker to semi-automatically finish the type checking work instead of working manually.

We organize our paper in the following way: section 2 introduces the related work on typing situation calculus and its variants; Section 3 presents the basic ideas on type systems and situation calculus in a straightforward way; Section 4 illustrates the primary ideas on how to type a lightweight core of the original situation calculus; Section 5 evaluates our typing mechanisms by type checking an existing piece of program in situation calculus and section 6 concludes this paper.

\section{Related Work}

Due to its powerful action-based reasoning ability, situation calculus is often chosen as the formalism to express other models and programming languages which are either too complex to understand and use, like Artificial Intelligence in games \cite{AIGame} and Planning Domain Definition Language (PDDL) \cite{PDDL}, or a little powerless to represent an entire complicated systems of different types, like Action Description Language (ADL) \cite{ADL}. In the literature employing situation calculus as a formal method to express the semantics in PDDL \cite{PDDLSC} and ADL \cite{ADLSC}, the authors have tried to introduce some typing mechanisms, which is only limited to add type element in syntax, and only applied to variables. Other significant terms, such as fluents and predicates, are still typeless. Moreover, in semantics and reasoning, typing mechanisms are hardly discussed in these papers, neither is type checking.

Yilan Gu et al. \cite{MSC} proposed a modified version of the situation calculus built using a two-variable fragment of the first-order logic extended with counting quantifiers. By introducing several additional groups of axiom to capture taxonomic reasoning and using similar regression operator in Raymond Reiter's work \cite{KiA}, the projection and executability problems are proved decidable although an initial knowledge base is incomplete and open. While their system concerns primarily on semantics of the new components proposed but rarely talks about typing on them, our well-typed version of situation calculus mentions typing mechanisms together with a modified situation calculus version in an all around way.

There are also some attempts on modifying situation calculus only based on a lightweight version of the original one. Gerhard Lakemeyer et al. \cite{MSC2} proposed a new logic dialect of situation calculus with the situation terms suppressed, namely, \includegraphics[height=0.3cm]{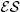}. That is, it is merely a similar formalism as a part of the current situation calculus. Moreover, in this paper, the authors consider how to map sentences between \includegraphics[height=0.3cm]{es.png} and situation calculus and try to prove \includegraphics[height=0.3cm]{es.png} is powerful enough to handle most cases as the situation calculus does, but mention little about how to type their new logic system as a fragment of situation calculus.

\section{Background Knowledge}

\subsection{Type Systems}
In the discipline of computer science, modern type systems are regarded as a formal mechanism originated from Alonzo Church's $\lambda$ calculus proposed in 1940 \cite{TS}. One possible definition of a type system is ``a tractable syntactic framework for classifying phrases according to the kinds of values they compute" \cite{TAPL}. By associating types with each computed value, a compiler can detect meaningless or invalid code written in a given programming language. For instance, the expression ``mix = 29 + ``Tan"" cannot get through type checking since a string cannot be added to a number.

There are many branches in type systems, such as inferred typing and manifest typing (implicit and explicit), and strong typing and weak typing. As for type checking, people can utilize dynamic type checking and static type checking, or a combination of both.

The primary and most obvious purpose of using type systems is to guarantee the correctness of programs, i.e., detect potential errors, while a well-typed system can further ensure the soundness (safety) of programs. The most important characteristics of a well-typed system are properties of preservation and progress. The former one makes sure a term can keep its type passed into the term that it is evaluated to, and the latter keeps reachability of a term: a typed term can either turn into a value or another related term, which means a well-typed term will not get stuck. In this paper, we plan to equip the current version of situation calculus with appropriate type system besides several original ones for variables. Thus, a program written in situation calculus can be easily type checked correct or not.

\subsection{Situation Calculus}
Situation Calculus \cite{SC} is a formal method based on First Order Logic and Set Theory in Mathematical Logic, with a strong ability of action-based reasoning. This formalism is considered as a dialect of logic programming language and mostly used in dynamic domain modeling. In situation calculus, the world is comprised of situations, actions and objects. The semantics of these three key components in situation calculus is given informally below.

A situation represents a possible world history, simply a sequence of actions, denoted by a first-order term. The constant $s_0$ is used to denote the initial situation, namely, the empty sequence of actions.

An action represents any possible change to the world, denoted by a function, for example, \emph{drop(A)}, \emph{clean(B)} and \emph{check\_in(ID)}.

An object represents an entity defined in the domain of a specific application, denoted by a first-order term, for example, \emph{x}, \emph{robot\_A} and \emph{table}.

Moreover, other than aforementioned three elements, there is another significant symbol used frequently in situation calculus, namely, fluents. A fluent represents a relation or a function whose truth values varies from one situation to the next, called relational fluent or functional fluent respectively.

Additionally, introduce two predefined binary symbols of fluents as follows:

Function symbol \emph{do} is defined as \emph{do: Action $\times$ Situation $\rightarrow$ Situation}, which maps an action \emph{a} and a situation \emph{s} to a new situation called successor situation, which results from performing the action \emph{a} in the situation \emph{s}. This successor situation is denoted as \emph{do(a, s)}.

Predicate symbol \emph{Poss} is defined as \emph{Poss: Action $\times$ Situation}. Similarly as above, \emph{Poss(a, s)} means it is possible to execute the action \emph{a} in the situation \emph{s}. Note that in the original situation calculus, there is no return value for \emph{Poss}. For consistency, in our well-typed system, we assign a \emph{unit} value for every \emph{Poss} predicate. In other words, \emph{Poss} is defined as \emph{Poss: Action $\times$ Situation $\rightarrow$ Unit} (Capital "U" indicates it is a type but not a value.).

As mentioned before, fluents are used to represent a term whose value varies according to the changing of situations. As a comparison, another symbol is defined to denote a term whose value does varies with situations, namely, predicate. For example, \emph{hunger\_status(person, time)} and \emph{weather\_condition(location, season)} are relational fluents while \emph{drop(person, object)} is a predicate, since in the first two fluents, the second arguments are actually situations, namely in situation calculus, s, and in the third term there is no specific situation, but only two objects, which means the value of this term will not change when situation changes.

\section{A Well-typed Mechanism in Situation Calculus}

\subsection{A Lightweight Situation Calculus}
The situation calculus we study and try to extend here is a lightweight version of its original form. Similarly as Featherweight Java (FJ), we only grab some core features in situation calculus and skip derivable forms to keep our ideas concise and efficient.

According to the language of situation calculus, we keep all the static domain element: situations, actions and objects, and the majority of functional elements like fluents \textbf{\emph{do}} and \textbf{\emph{poss}}, and all the predicates. The components we ignore are those that either can derive from other elements or similarly be expressed by others. For instance, the ordering predicate $\sqsubseteq$, which defines an ordering relation on situations, can be expressed implicitly by the return value of other fluents and predicates. Like the expression $s'$ $\sqsubseteq$ $s$ which denotes that $s'$ is a proper subsequence of $s$, $s'$ could be replaced with a fluent or predicate which leads $s$ to $s'$, say, \emph{\textbf{do}(findajob(person:Object, job:Object), s:Situation)}. 

Likewise, we replace countably infinitely many predicate symbols with arity \emph{n}, $(action \cup object)^n$ with $\overline{t}$, which is a shorthand of a sequence of terms $t_1, t_2, \ldots, t_n~(n \geq 1)$.

\subsection{Handy Typing Mechanism}
In the original situation calculus, several elements such as quantifiers are typed \cite{KiA}. The handy typed elements are described formally as follows:

A typed notion \emph{$\tau$(x)} is used to denote \emph{x} associated with a finite set of all possible types:

$\tau(x) \stackrel{\mathrm{def}}{=} x : T_1 \vee x : T_2 \vee \ldots \vee x : T_n$, where $T_1, T_2, \ldots, T_n$ are types of terms.

Moreover, typed quantifiers are given by virtue of:

$(\forall x : \tau)\phi(x) \stackrel{\mathrm{def}}{=} (\forall x).\tau(x) \supset \phi(x)$,

$(\exists x : \tau)\phi(x) \stackrel{\mathrm{def}}{=} (\exists x).\tau(x) \wedge \phi(x)$.

Thus, expressions that contain such typed quantifiers could be rewritten as sequences of conjunctions and disjunctions:

$(\forall x : \tau)\phi(x) \equiv \phi(T_1) \vee \phi(T_2) \vee \ldots \vee \phi(T_n)$,

$(\exists x : \tau)\phi(x) \equiv \phi(T_1) \wedge \phi(T_2) \wedge \ldots \wedge \phi(T_n)$.

\subsection{A New Type System in the Lightweight Situation Calculus}
Although the original version of situation calculus equips some components with corresponding types and semantics, it is not enough to do type checking based on these definitions. We proposed a new well-typed system to enable potential task of type checking in a convenient way.

\vspace{2mm}
\noindent\textbf{Syntactic Forms}
\vspace{1mm}

$t ::= \ldots \hfill \textbf{terms:}$

$~~~~~~~~~x \hfill variable$

$~~~~~~~~~\forall x \hfill universal~quantified~variable$

$~~~~~~~~~\exists x \hfill existential~quantified~variable$

$~~~~~~~~~\neg t \hfill negative~term$

$~~~~~~~~~t_1 \supset t_2 \hfill subset~logical~connection$

$~~~~~~~~~t_1 \wedge t_2 \hfill conjunction~logical~connection$

$~~~~~~~~~t_1 \vee t_2 \hfill disjunction~logical~connection$

$~~~~~~~~~\overline{t} \hfill term~sequence$

$bt ::= \ldots \hfill \textbf{behavioral terms:}$

$~~~~~~~~~\neg bt \hfill negative~behavioral~term$

$~~~~~~~~~r(\overline{t}, s) \hfill relational~fluent$

$~~~~~~~~~f(\overline{t}) \hfill predicate$


$~~~~~~~~~\textbf{\emph{do}}(bt, s) \hfill functional~fluent$

$~~~~~~~~~\textbf{\emph{poss}}(bt, s) \hfill predicate~fluent$

$v ::= \ldots \hfill \textbf{values:}$

$~~~~~~~~~unit \hfill poss~predicate~value$

$~~~~~~~~~true \hfill true~boolean~value$

$~~~~~~~~~false \hfill false~boolean~value$

$T ::= \ldots \hfill \textbf{types:}$

$~~~~~~~~~Unit \hfill type~of~predicate~fluent$

$~~~~~~~~~Bool \hfill type~of~booleans$

$~~~~~~~~~Situation \hfill type~of~behavioral~terms$

$~~~~~~~~~Action \hfill type~of~behavioral~terms$

$~~~~~~~~~Object \hfill type~of~terms$

\vspace{2mm}
\noindent\textbf{Semantics}
\vspace{1mm}

Given a world \emph{w} comprised of situations, actions and objects and a set \emph{L(w)} of all instances defined in \emph{w}, if a term \emph{t} holds in \emph{w}, we write \emph{w} $\models$ \emph{t}. Given a set of situations $S = \{s_0, s_1, \ldots, s_n\}~(n \geq 0)$, we have:

\begin{tabular}{ll}
$w \models x$ & $\Leftrightarrow~~~~x \in L(w)$ 
\\
$w \models \forall x$ & $\Leftrightarrow~~~~\forall s_i \in S, w \models x$
\\
$w \models \exists x$ & $\Leftrightarrow~~~~\exists s_i \in S, w \models x$
\\
$w \models \neg x$ & $\Leftrightarrow~~~~w \not\models x$
\\
$w \models t_1 \supset t_2$ & $\Leftrightarrow~~~~w \models t_1 \Rightarrow w \models t_2$
\\
$w \models t_1 \wedge t_2$ & $\Leftrightarrow~~~~w \models t_1~\mathrm{and}~w \models t_2$
\\
$w \models t_1 \vee t_2$ & $\Leftrightarrow~~~~w \models t_1~\mathrm{or}~w \models t_2$
\\
$w \models \overline{t}$ & $\Leftrightarrow~~~~w \models t_1, w \models t_2, \ldots, w \models t_n$
\\
$w \models \neg bt$ & $\Leftrightarrow~~~~w \not\models bt$
\\
$w \models r(\overline{t}, s)$ & $\Leftrightarrow~~~~w \models \overline{t}~\mathrm{and}~w \models s~in~r$
\\
$w \models f(\overline{t})$ & $\Leftrightarrow~~~~w \models \overline{t}~in~f$
\\
$w \models \textbf{\emph{do}}(bt, s)$ & $\Leftrightarrow~~~~\exists s_i \in S, bt~holds~in~s_i$
\\
$w \models \textbf{\emph{poss}}(bt, s)$ & $\Leftrightarrow~~~~\exists s_i \in S, w \models \big (s_i \supset \textbf{\emph{do}}(bt, s_i)\big )$
\end{tabular}

\vspace{2mm}
\noindent\textbf{Evaluation Rules} \hfill \textsf{\textbf{t $\rightarrow$ t'}}
\vspace{5mm}

\large $\frac{(x)bt~\rightarrow~(x')bt}{(\forall x)bt~\rightarrow~(\forall x')bt}$ \hfill \normalsize \textsc{E-Unv}
\vspace{1mm}

\large $\frac{(x)bt~\rightarrow~(x')bt}{(\exists x)bt~\rightarrow~(\exists x')bt}$ \hfill \normalsize \textsc{E-Est}
\vspace{1mm}

\large $\frac{t~\rightarrow~t'}{\neg t~\rightarrow~\neg t'}$, $\frac{bt~\rightarrow~bt'}{\neg bt~\rightarrow~\neg bt'}$ \hfill \normalsize \textsc{E-Neg}
\vspace{1mm}

\large $\frac{t_1~\rightarrow~t_1'}{t_1~\supset~t_2~\rightarrow~t_1'~\supset~t_2}$ \hfill \normalsize \textsc{E-Spt}
\vspace{1mm}

\large $\frac{t_1~\rightarrow~t_1'}{t_1~\wedge~t_2~\rightarrow~t_1'~\wedge~t_2}$ \hfill \normalsize \textsc{E-Conj}
\vspace{1mm}

\large $\frac{t_1~\rightarrow~t_1'}{t_1~\vee~t_2~\rightarrow~t_1'~\vee~t_2}$ \hfill \normalsize \textsc{E-Disj}
\vspace{1mm}

\large $\frac{t_1~\rightarrow~t_1'}{t_1,~t_2,~\ldots,~t_n~\rightarrow~t_1',~t_2,~\ldots,~t_n}$ \hfill \normalsize \textsc{E-Seq}\vspace{2mm}

$\textbf{\emph{do}}(bt, s) \rightarrow [s \mapsto s']bt$ \hfill \textsc{E-Do}\vspace{2mm}

$\textbf{\emph{poss}}(bt, s) \rightarrow s \supset [s \mapsto s']bt$ \hfill \textsc{E-Poss}

\normalsize

\vspace{2mm}
\noindent\textbf{Typing Rules} \hfill \textsf{\textbf{W $\vdash$ t : T}}
\vspace{1mm}

Here we continue to use \emph{W} (rather than the lower case \emph{w} used in semantics) instead of conventional $\Gamma$ to denote a typing context. Formally, we have:\vspace{3mm}

$\emph{W}~\vdash~true : Bool$ \hfill \textsc{T-True}
\vspace{3mm}

$\emph{W}~\vdash~false : Bool$ \hfill \textsc{T-False}
\vspace{3mm}

\large $\frac{x~:~T~\in~\normalsize \emph{W}}{\normalsize \emph{W}~\vdash~x~:~T}$ \hfill \normalsize \textsc{T-Var}
\vspace{3mm}

\colorbox{lightgray}{\large $\frac{\forall r(x~:~T,~\overline{t}-x,~s)~\in~\normalsize \emph{W}}{\normalsize \emph{W}~\vdash~(\forall x~:~T)~r(\overline{t},~s)}$\hspace{4in}~~~~~~\normalsize \textsc{T-Unv1}} 
\vspace{3mm}

\colorbox{lightgray}{\large $\frac{\exists r(x~:~T,~\overline{t}-x,~s)~\in~\normalsize \emph{W}}{\normalsize \emph{W}~\vdash~(\exists x~:~T)~r(\overline{t},~s)}$\hspace{4in}\hspace{0.2mm}~~~~~~~\normalsize \textsc{T-Est1}} 
\vspace{3mm}

\colorbox{lightgray}{\large $\frac{\forall f(x~:~T,~\overline{t}-x)~\in~\normalsize \emph{W}}{\normalsize \emph{W}~\vdash~(\forall x~:~T)~f(\overline{t})}$\hspace{4in}\hspace{0.5mm}~~~~~~~~\normalsize \textsc{T-Unv2}} 
\vspace{3mm}

\colorbox{lightgray}{\large $\frac{\exists f(x~:~T,~\overline{t}-x)~\in~\normalsize \emph{W}}{\normalsize \emph{W}~\vdash~(\exists x~:~T)~f(\overline{t})}$\hspace{4in}\hspace{0.7mm}~~~~~~~~~\normalsize \textsc{T-Est2}}\vspace{1mm} 
\vspace{3mm}

\large $\frac{\normalsize \emph{W}~\vdash~t~:~T}{\normalsize \emph{W}~\vdash~\neg t~:~T}$, $\frac{\normalsize \emph{W}~\vdash~bt~:~T}{\normalsize \emph{W}~\vdash~\neg bt~:~T}$ \hfill \normalsize \textsc{T-Neg}
\vspace{3mm}

\large $\frac{\normalsize \emph{W}~\vdash~(t_1~:~T_1)~\supset~(t_2~:~T_2)}{\normalsize \emph{W}~\vdash~(\forall x~\in~t_1)~x~:~T_1~\supset~(\forall y~\in~t_2)~y~:~T_2}$ \hfill \normalsize \textsc{T-Spt}
\vspace{3mm}

\large $\frac{\normalsize \emph{W}~\vdash~(t_1~:~T_1)~\wedge~(t_2~:~T_2)}{\normalsize \emph{W}~\vdash~(\forall x~\in~t_1)~x~:~T_1~\wedge~(\forall y~\in~t_2)~y~:~T_2}$ \hfill \normalsize \textsc{T-Conj}
\vspace{3mm}

\large $\frac{\normalsize \emph{W}~\vdash~(t_1~:~T_1)~\vee~(t_2~:~T_2)}{\normalsize \emph{W}~\vdash~(\forall x~\in~t_1)~x~:~T_1~\vee~(\forall y~\in~t_2)~y~:~T_2}$ \hfill \normalsize \textsc{T-Disj}
\vspace{3mm}

\large $\frac{\normalsize \emph{W}~\vdash~(t_1~:~T_1),~(t_2~:~T_2),~\ldots,~(t_n~:~T_n)}{\normalsize \emph{W}~\vdash~(\forall x~\in~t_1)~x~:~T_1,~\ldots,~(\forall z~\in~t_n)~z~:~T_n}$ \hfill \normalsize \textsc{T-Seq}
\vspace{3mm}

\large $\frac{\small \emph{W}~\vdash~\scriptsize r~:~Object\rightarrow Situation\rightarrow Situation,~\scriptsize~ \overline{t}~:~Object,~\scriptsize~s~:~Situation}{\normalsize \emph{W}~\vdash~r(\overline{t},~s)~:~Situation}$ \hfill \normalsize \textsc{T-RelFlt} 
\vspace{3mm}

\large $\frac{\normalsize \emph{W}~\vdash~f~:~Object\rightarrow Action~~\normalsize \emph{W}\vdash \overline{t}~:~Object}{\normalsize \emph{W}~\vdash~f(\overline{t})~:~Action}$ \hfill \normalsize \textsc{T-FunFlt}
\vspace{3mm}

\large $\frac{\normalsize \emph{W},~bt~:~Action~\vdash~s~:~Situation}{\normalsize \emph{W}~\vdash~\textbf{\emph{do}}(bt,~s)~:~Situation}$ \hfill \normalsize \textsc{T-Do}
\vspace{3mm}

\large $\frac{\normalsize \emph{W},~bt~:~Action~\vdash~s~:~Situation}{\normalsize \emph{W}~\vdash~\textbf{\emph{poss}}(bt,~s)~:~Unit}$ \hfill \normalsize \textsc{T-Poss}


\begin{center}
\begin{tabular}{|p{1.8cm}|p{13cm}|} 
\hline
\textbf{Item} & \textbf{Description} \\
\hline
\textsc{E-Unv} & If one term $t'$ occurred in a given behavioral term $bt$ derives from another term $t$ also in $bt$, then all such terms $t'$ in $bt$ are also derivable from all such terms $t$ in $bt$. \\
\hline
\textsc{E-Est} & If one term $t'$ occurred in a given behavioral term $bt$ derives from another term $t$ also in $bt$, then there exists such a term $t'$ in $bt$ that is derivable from such a term $t$ in $bt$. \\
\hline
\textsc{E-Neg} & If one term/behavioral term $t'$/$bt'$ derives from another term $t$/$bt$, then \texttt{not} $t'$/$bt'$ also derives from \texttt{not} $t$/$bt$. \\
\hline
\textsc{E-Spt} & If one term $t'$ derives from another term $t$, then this also holds in \texttt{superset} operation. \\
\hline
\textsc{E-Conj} & If one term $t'$ derives from another term $t$, then this also holds in \texttt{conjunction}. \\
\hline
\textsc{E-Disj} & If one term $t'$ derives from another term $t$, then this also holds in \texttt{disjunction}. \\
\hline
\textsc{E-Seq} & If one term $t'$ derives from another term $t$, then this relationship holds if $t'$ and $t$ are in a sequence of terms, respectively. \\
\hline
\textsc{E-Do} & In a specific situation $s$, behavioral term $bt$ gets executed means situation $s$ transits to its successor situation $s'$ while doing $bt$. \\
\hline
\textsc{E-Poss} & In a specific situation $s$, behavioral term $bt$ is possible means current situation $s$ is a superset of its successor situation $s'$. \\
\hline
\textsc{T-True} & As a \texttt{Bool} type value, \texttt{true} is within the typing map \emph{W}. \\
\hline
\textsc{T-False} & As a \texttt{Bool} type value, \texttt{false} is within the typing map \emph{W}. \\
\hline
\textsc{T-Var} & If a variable $x$ with type $T$ is within the typing map $W$, then $x:T$ derives from $W$. \\
\hline
\textsc{T-Unv1} & If all relational fluents $r$ that have an argument $x$ with type $T$ hold, then all occurrence of $x$ in $r$ must have a type $T$. \\
\hline
\textsc{T-Est1} & If there exists one relational fluent $r$ that has an argument $x$ with type $T$ hold, then there must be one occurrence of $x$ in $r$ with a type $T$. \\
\hline
\textsc{T-Unv2} & If all functional fluents $f$ that have an argument $x$ with type $T$ hold, then all occurrence of $x$ in $r$ must have a type $T$. \\
\hline
\textsc{T-Est2} & If there exists one functional fluent $f$ that has an argument $x$ with type $T$ hold, then there must be one occurrence of $x$ in $r$ with a type $T$. \\
\hline
\textsc{T-Neg} & If one term/behavioral term $t'$/$bt'$ with a type $T$ derives from the typing map $W$, then \texttt{not} $t'$/$bt'$ also derives from $W$ with the same type $T$. \\
\hline
\textsc{T-Spt} & If $t_1$ with a type $T_1$ as a superset of $t_2$ with a type $T_2$ derives from the typing map $W$, then in $W$, all subterms $x$ of $t_1$, $y$ of $t_2$ also have types $T_1$, $T_2$ respectively, and \texttt{superset} relationship still holds. \\
\hline
\textsc{T-Conj} & If the conjunction of $t_1$ with a type $T_1$ and $t_2$ with a type $T_2$ derives from the typing map $W$, then all subterms of $t_1$, $t_2$ also have types $T_1$, $T_2$, satisfying \texttt{conjunction}. \\
\hline
\textsc{T-Disj} & If the disjunction of $t_1$ with a type $T_1$ and $t_2$ with a type $T_2$ derives from the typing map $W$, then all subterms of $t_1$, $t_2$ also have types $T_1$, $T_2$, satisfying \texttt{disjunction}. \\
\hline
\textsc{T-Seq} & If a sequence of terms with its own types derives from the typing map $W$, then in $W$, all subterms of every term have the type their parent has. \\
\hline
\textsc{T-RelFlt} & Straightforward typing relationship of first-order logic. \\
\hline
\textsc{T-FunFlt} & Straightforward typing relationship of first-order logic. \\
\hline
\textsc{T-Do} & Straightforward typing relationship of first-order logic. \\
\hline
\textsc{T-Poss} & Straightforward typing relationship of first-order logic. \\
\hline
\end{tabular}\\\vspace{1em}
Table 1: A Directory of All Evaluation and Typing Rules in the Type System of the Lightweight Situation Calculus
\end{center}


\textbf{Note}:

1. $\overline{t}$ is a shorthand of a sequence of terms $t_1, t_2, \ldots, t_n~(n \geq 1)$. Hence $\overline{t}$ cannot possibly be empty.

2. The type \emph{Unit} is defined as the type of value \emph{unit}, where $unit \stackrel{\mathrm{def}}{=} \{t|(\forall x\in t)~x:Bool \vee Situation \vee Action \vee Object\}$, which means all elements in a \emph{unit} should have the same type. 

3. $[s \mapsto s']bt$ in the computation rules \textsc{E-Do} and \textsc{E-Poss} means ``the successor situation $s'$ that results from executing the behavioral term $bt$ in the situation $s$." See the items for \textsc{E-Do} and \textsc{E-Poss} in Table 1.

4. The shadowed typing rules are adapted from the handy typing mechanism for quantifiers in current version of situation calculus, which is discussed in section 4.1.

5. For simplicity, the detailed explanation is not given for typing rules \textsc{T-RelFlt}, \textsc{T-FunFlt}, \textsc{T-Do} and \textsc{T-Poss}.

\section{Evaluation}

\subsection{Case Description}
Let us consider the following scenario: In face of an object \emph{x} on the floor, say a vase, there is a robot \emph{r} who wants to pick up this vase and paints it with some color, namely \emph{c}. In situation calculus, we can describe this scenario using three statements:

In a given situation \emph{s} that, say, there is a robot \emph{r} and a vase \emph{x} ready for situations later on, if the robot \emph{r} then picked up the vase \emph{x} and dropped it without holding it firmly, which made the vase became broken, then the vase must be a fragile object:

$fragile(x, s) \supset broken(x, \textbf{\emph{do}}(drop(r, x), s))$ \hfill (1)

If the robot successfully picked up the vase \emph{x} and tried to paint it with one color \emph{c}, holding it firmly, the vase would turn out to be in the color \emph{c}:

$color(x, \textbf{\emph{do}}(paint(x, c), s)) = c$ \hfill (2)

Finally, let us consider the conditions on which it is possible for the robot \emph{r} to pick up the vase \emph{x} without any external help. The conditions should be a combination of three: the robot \emph{r} is not holding any other object \emph{z}, it is next to \emph{x}, and \emph{x} is not heavy:

$\textbf{\emph{poss}}(pickup(r, x), s) \supset [(\forall z)\neg holding(r, z, s)] \wedge \neg heavy(x) \wedge nextTo(r, x, s)$ \hfill (3)

\subsection{Results and Analysis of Type Checking}
Now let us do the type checking on the aforementioned three statements that represent a scenario in which a robot \emph{r} wants to pick up a vase \emph{x} by itself and paints it with some fancy color \emph{c}. On the basis of the type system defined in section 4.2, if all the typing goes through and does not get stuck, the program written in situation calculus will be regarded as well-typed.

First, we need to add predefined types for programs written in the original situation calculus by virtue of our new type system. Hence we have:

$fragile(x:Object, s:Situation) \supset broken(x:Object, \textbf{\emph{do}}(drop(r:Object, x:Object), s:Situation))$

 \hfill $(1)^\prime$ 

$color(x:Object, \textbf{\emph{do}}(paint(x:Object, c:Object), s:Situation)) = c:Object$ \hfill $(2)^\prime$

$\textbf{\emph{poss}}(pickup(r:Object, x:Object), s:Situation) \supset [(\forall z:Object)\neg holding(r:Object, z:Object, s:Situation)] \wedge \neg heavy(x:Object) \wedge nextTo(r:Object, x:Object, s:Situation)$ \hfill $(3)^\prime$

And then we know the world $w \equiv \{x, s, r, c, z\}$ and $W \equiv \{x: Object, s: Situation, r: Object, c: Object, z:Object\}$.

Now, let us do typing derivation statement by statement. For $(1)^\prime$, We notice that \textsc{T-Spt} cannot be applied since $(1)^\prime$ is of a superset relationship between behavioral terms while \textsc{T-Spt} is for regular terms. Thus, we turn to prove that the type of the left hand side of ``$\supset$'' is the same as that of the right hand side.

For typesetting simplicity, we omit ``\emph{W}$\vdash$'', return types and the final step of \textsc{T-Var}, and abbreviate ``\emph{Object}'', ``\emph{Situation}'' and ``\emph{Action}'' to ``\emph{Obj}'', ``\emph{Stn}'' and ``\emph{Atn}'', respectively, in the following typing derivation.

Left hand side of ``$\supset$'' in $(1)^\prime$:

\vspace{-3mm}
$$\frac{fragile:Obj\rightarrow Stn\rightarrow Stn,~x:Obj,~s:Stn}{fragile(x,~s)}\textsc{T-RelFlt}$$ 
\vspace{-3mm}

Right hand side of ``$\supset$'' in $(1)^\prime$:

\vspace{-3mm}
$$\frac{drop:Obj\rightarrow Atn,~r:Obj,~x:Obj,~s:Stn,~broken:Obj\rightarrow Stn\rightarrow Stn}{drop(r:Obj, x:Obj),~s:Stn,~broken:Obj\rightarrow Stn\rightarrow Stn}\textsc{T-FunFlt}$$\vspace{-6mm}
$$\hspace{-8mm}\frac{~}{~~~~~~~\textbf{\emph{do}}(drop(r:Obj, x:Obj),~s:Stn),~broken:Obj\rightarrow Stn\rightarrow Stn~~~~~~~~}\textsc{T-Do}$$\vspace{-6mm}
$$\frac{~}{~~~~~~~~~~~~~~~~broken(x:Obj,~\textbf{\emph{do}}(drop(r:Obj, x:Obj),~s:Stn))~~~~~~~~~~~~~~~~}\textsc{T-RelFlt}$$
\vspace{-3mm}

According to this typing derivation, we know that both types of the left hand side and right hand side are the same one: \emph{Situation}. So this situation calculus statement is proved to be well-typed.

For $(2)^\prime$, we have the similar form of typing derivation:

Left hand side of ``='' in $(2)^\prime$:

\vspace{-3mm}
$$\frac{\hspace{0.8mm}paint:Obj\rightarrow Atn,~x:Obj,c:Obj,~s:Stn,~color:Obj\rightarrow Stn\rightarrow Stn}{paint(x:Obj, c:Obj),~s:Stn,~color:Obj\rightarrow Stn\rightarrow Stn}\textsc{T-FunFlt}$$\vspace{-6mm}
$$\hspace{-8mm}\frac{~}{~~~~~~~\textbf{\emph{do}}(paint(x:Obj, c:Obj),~s:Stn),~color:Obj\rightarrow Stn\rightarrow Stn~~~~~~~}\textsc{T-Do}$$\vspace{-6mm}
$$\frac{~}{~~~~~~~~~~~~~~~color(x:Obj,~\textbf{\emph{do}}(paint(x:Obj, c:Obj),~s:Stn))~~~~~~~~~~~~~~~~}\textsc{T-RelFlt}$$
\vspace{-3mm}

Right hand side of ``='' in $(2)^\prime$:

\emph{c: Object}

According to this typing derivation, we find that the type of the left hand side is \emph{Situation}, while the right hand side has a type: \emph{Object}, which obviously leads to a mismatch. So this situation calculus statement is proved to be not well-typed. In fact, whatever type of \emph{c} will bring about stuck terms or mismatches. Anyway, it can still be fixed. A possible correction is to change the right hand side to \emph{inColor(c: Object, s: Situation)}, i.e., to replace \emph{c} with a corresponding relational fluent to match the type of the left hand side.

Let us check the final sample statement similarly as we did previously:

Left hand side of ``$\supset$'' in $(3)^\prime$:

\vspace{-3mm}
$$\frac{pickup:Obj\rightarrow Obj\rightarrow Atn,~r:Obj,~x:Obj,~s:Stn}{pickup(r:Obj,~x:Obj),~s:Stn}\textsc{T-FunFlt}$$\vspace{-6mm}
$$\hspace{-5mm}\frac{~}{~~~~~~~~~~~~\textbf{\emph{poss}}(pickup(r:Obj,~x:Obj),~ s:Stn)~~~~~~~~~~~}\textsc{T-Poss}$$
\vspace{-3mm}

When coming to the right hand side of ``$\supset$'' in $(3)^\prime$, we notice that \textsc{T-Conj} cannot be applied since the right hand side of $(3)^\prime$ is of a conjunctive relationship between behavioral terms while \textsc{T-Conj} is for regular terms. Thus, we turn to check whether types of each part of the conjunction are the same. If so, the final type should be \emph{Unit} according to its definition.

\vspace{-2mm}
$$\frac{holding:Obj\rightarrow Obj\rightarrow Stn\rightarrow Stn,~r:Obj,~z:Obj,~s:Stn}{holding(r:Obj,~z:Obj,~s:Stn)}\textsc{T-RelFlt}$$\vspace{-6mm}
$$\hspace{-6mm}\frac{~}{~~~~~~~~~~~~~~~~~~~\forall \neg holding(r:Obj,~z:Obj,~ s:Stn)~~~~~~~~~~~~~~~~~~~~}\textsc{T-Neg}$$\vspace{-6mm}
$$\hspace{-3mm}\frac{~}{~~~~~~~~~~~~(\forall z:Obj)\neg holding(r:Obj,~z:Obj,~s:Stn)~~~~~~~~~~~~~}\textsc{T-Unv1}$$
\vspace{-3mm}

So we find that $(\forall z)\neg holding(r:Object, z:Object, s:Situation)$ has a type \emph{Situation}. Similarly, we can derive $\neg heavy(x:Object)$ to be of type \emph{Action} by taking two typing derivation steps and $nextTo(r:Object, x:Object, s:Situation)$ to be of type \emph{Situation} by taking one step. By definition, the type of the right hand side of ``$\supset$'' in $(3)^\prime$ is not \emph{Unit} since not all the subterms of the conjunction have the same type. Therefore, we can claim that $(3)^\prime$ is not well-typed as well. This time we can fix it intuitively by simply changing the functional fluent $\neg heavy(x:Object)$ into a relational fluent $\neg heavy(x:Object, s:Situation)$.

\subsection{Implementation in OCaml}
In the last section, types of every term and behavioral term written in a modified lightweight situation calculus with our type system are checked for consistency theoretically. Now we plan to implement a type checker in OCaml which does the same job as we do manually, that is, all the type checking work would be fulfilled by a type checker semi-automatically and efficiently, which can give a great hand for those who are doing tedious type checking alone.

One piece of sample code in OCaml which typechecks situation calculus statement $(1)^\prime$ as described in 5.2 is shown below:
\\

\noindent \verb"(* Types Definition *)"\\
\verb"# type unit = Unit of unit;;"\\ 
\verb"# type bool = Bool of bool;;"\\
\verb"# type stn  = Situation;;"\\
\verb"# type atn  = Action;;"\\
\verb"# type obj  = Object;;"\\

\noindent \verb"(* T-RelFlt *)"\\
\verb"# let r t s = "\\
\verb"      match t with"\\
\verb"          Object -> match s with"\\
\verb"                        Situation -> Situation;;"\\

\noindent \verb"(* T-FunFlt *)"\\
\verb"# let f t1 t2 ="\\
\verb"      match t1 with"\\
\verb"          Object -> match t2 with"\\
\verb"                        Object -> Action;;"\\

\noindent \verb"(* T-Do *)"\\
\verb"# let does bt s ="\\
\verb"      match bt with"\\
\verb"          Action -> match s with"\\
\verb"                        Situation -> Situation;;"\\

\noindent \verb"(* Left Hand Side *)"\\
\verb"# let x = Object"\\
\verb"  and s = Situation"\\
\verb"  and fragile = r;;"\\
\verb"val x : obj = Object"\\
\verb"val s : stn = Situation"\\
\verb"val fragile : obj -> stn -> stn = <fun>"\\
\verb"# fragile (x:obj) (s:stn);;"\\
\verb"- : stn = Situation"\\

\noindent \verb"(* Right Hand Side *)"\\
\verb"# let b = Object"\\
\verb"  and drop = f"\\
\verb"  and broken = r;;"\\
\verb"val b : obj = Object"\\
\verb"val drop : obj -> obj -> atn = <fun>"\\
\verb"val broken : obj -> stn -> stn = <fun>"\\
\verb"# broken (x:obj) (does (drop (b:obj) (x:obj)) (s:stn));;"\\
\verb"- : stn = Situation"\\

\noindent \verb"(* This statement is proved to be well-typed *)"\\

In the OCaml code above, we firstly defined the types in our type system, and then implemented the \textsc{T-RelFlt}, \textsc{T-FunFlt} and \textsc{T-Do} typing rules. Finally some necessary variables, two relational fluents \emph{fragile} and \emph{broken}, and a funtional fluent \emph{drop} are declared. As a type checking process, these fluents \emph{fragile}, \emph{broken} and \emph{drop} are invoked with inputs of pre-defined variables to show the typing relationship among them, and the final types calculated for the left hand side and right hand side indicate whether this statement is well-typed.

In this way, all statements that we typechecked manually just now can be dumped into this type checker for semi-automatical type checking. Due to some limitation of typing rules in our system, we do need some additional manual work occasionally. For instance, we need to check by ourselves that whether the types of the left hand side and right hand side of a symbol "$\supset$" are the same. Anyway, the type checker indeed facilitate our process of deciding whether a situation calculus program is well-typed or not.

\section{Conclusions}
Type systems have been proposed to guarantee the soundness of program types by rigid typing mechanisms. As a popular formal language widely used in Artificial Intelligence related fields, situation calculus itself has insufficient methods to support a complete and robust type system, with a rudimentary typing mechanism: only typed quantifiers for variables. It is obviously not enough for type checking the current situation calculus programs. By virtue of our newly-introduced type system for a lightweight situation calculus which keep the core of the current one, we can easily do basic type checking for existing situation calculus programs which are referred a lot in various study of situation calculus. We also implemented the theoretical type system in OCaml as a type checker to substantiate our ideas. With the help of this type checker, precedent manual type checking work can be greatly automated for a better performance. As for the programs checked out to be ill-typed, we provide corresponding ways for correcting them into well-typed forms.\\\vspace{1em}

\noindent \Large \textbf{Acknowledgements}\vspace{3mm}\\\normalsize
The author would like to thank all anonymous reviewers for their generous and constructive directives and comments on this paper.

\nocite{*}
\bibliographystyle{eptcs}
\bibliography{generic}
\end{document}